\documentclass[showpacs,amsmath,pra]{revtex4}

\usepackage{graphicx}
\usepackage{dcolumn}
\usepackage{bm}

\begin{document}

\title{{\LARGE Quantum exam}}
\author{\bf Nguyen Ba An}
\email{nbaan@kias.re.kr}
\affiliation{School of Computational Sciences, Korea Institute for Advanced
Study, 207-43 Cheongryangni 2-dong, Dongdaemun-gu, Seoul 130-722, Republic of Korea}

\begin{abstract}
Absolutely and asymptotically secure protocols for organizing an exam in a
quantum way are proposed basing judiciously on multipartite entanglement. The protocols
are shown to stand against common types of eavesdropping attack.
\end{abstract}
\pacs{03.67.Hk, 03.65.Ud, 03.67.Dd}
\maketitle

\noindent \textbf{1. Introduction}

Simultaneous distance-independent correlation between different systems
called entanglement \cite{r1} is the most characteristic trait that sharply
distinguishes between quantum and classical worlds. At present entanglement
between two systems, i.e. bipartite entanglement, is quite well understood,
but that between more than two systems, i.e. multipartite entanglement,
remains still far from being satisfactorily known. In spite of that,
multipartite entanglement has proven to play a superior role in recently
emerging fields of quantum information processing and quantum computing
since it exhibits a much richer structure than bipartite entanglement.
Motivation for studying multipartite entanglement arises from many reasons
some of which are listed now. First, multipartite entanglement provides a
unique means to check the Einstein locality without invoking statistical
arguments \cite{r2}, contrary to the case of Bell inequalities using bipartite
entanglement. Second, multipartite entanglement serves as a key ingredient
for quantum computing to achieve an exponential speedup over classical
computation \cite{r3}. Third, multipartite entanglement is central to
quantum error correction \cite{r4} where it is used to encode states, to
detect errors and, eventually, to allow fault-tolerant quantum computation 
\cite{r5}, Fourth, multipartite entanglement helps to better characterize
the critical behavior of different many-body quantum systems giving rise to
a unified treatment of the quantum phase transitions \cite{r6}. Fifth,
multipartite entanglement is crucial also in condensed matter phenomena and
might solve some unresolved problems such as high-T superconductivity \cite
{r7}. Sixth, multipartite entanglement is recognized as a unreplaceable or
efficient resource to perform tasks involving a large number of parties such
as network teleportation \cite{r8}, quantum cryptography \cite{r9}, quantum
secret sharing \cite{r10}, remote entangling \cite{r11}, quantum
(tele)cloning \cite{r12}, quantum Byzantine agreement \cite{r13}, etc.
Finally, multipartite entanglement is conjectured to yield a wealth of
fascinating and unexplored physics \cite{r14}. Current research in
multipartite entanglement is progressing along two directions in parallel.
One direction deals with problems such as how to classify \cite{r15},
quantify \cite{r16}, generate/control/distill \cite{r17} and witness \cite
{r18} multipartite entanglement. The other direction proceeds to advance
various applications exploiting the nonclassical multiway correlation 
inherent in multipartite entanglement \cite{r8,r9,r10,r11,r12,r13}. 
Our work here
belongs to the second direction mentioned above. Namely, we propose
protocols to organize the so-called quantum exam which will be specified in
the next section. To meet the necessary confidentiality of the exam we use
suitable multipartite GHZ entangled states \cite{r2} as the quantum channel.
We consider two scenarios. One scenario is absolutely secure provided that
the participants share a prior proper multipartite entanglement. The other
scenario can be performed directly without any nonlocal quantum arrangements in the
past but it is only asymptotically secure. Both the scenarios are shown to
stand against commonly utilized eavesdropping attacks.

\vskip 0.5cm

\noindent \textbf{2. Quantum exam}

Exploiting the superdense coding feature possessed in bipartite entanglement
we have recently proposed a quantum dialogue scheme \cite{r19} (see also 
\cite{r20}) allowing two legitimate parties to securely carry out their
conversation. In this work multipartite entanglement will be judiciously exploited to do
a more sophisticated task. Suppose that a teacher Alice wishes to organize
an important exam with her remotely separate students Bob $1$, Bob $2$,
..... and Bob $N.$ Alice gives her problem to all Bobs and, after some
predetermined period of time, asks each Bob to return a solution
independently. Alice's problem should be kept confidential from any
outsiders. The solution of a Bob should be accessible only to Alice but not
to anyone else including the $N-1$ remaining Bobs. Such confidentiality
constraints cannot be maintained even when Alice and Bobs are connected by
authentic classical channels because any classical communication could be
eavesdropped perfectly without a track left behind. However, combined with
appropriate quantum channels such an exam is accomplishable. We call it
quantum exam, i.e. an exam organized in a quantum way to guarantee the
required secrecy.

Let Alice's problem is a binary string 
\begin{equation}
Q=\{q_{m}\}  \label{Q}
\end{equation}
and the solution of a Bob is another string 
\begin{equation}
R_{n}=\{r_{nm}\}  \label{R}
\end{equation}
where $n=1,$ $2,$ $...,$ $N$ labels the Bob while $q_{m},$ $r_{nm}\in
\{0,1\} $ with $m=1,$ $2,$ $3,...$ denote a secret bit of Alice and a Bob.

\vskip 0.5cm

\noindent \textbf{2.1. Absolutely secure protocol}

An exam consists of two stages. In the first stage Alice gives a problem to
Bobs and in the second stage she collects Bobs' solutions.
\vskip 0.5cm
\noindent \textbf{The problem-giving process}

To securely transfer the problem from Alice to Bobs the following steps are
to be proceeded. 

\begin{enumerate}
\item[a1)]  Alice and Bobs share beforehand a large number of ordered
identical $(N+1)$-partite GHZ states in the form 
\begin{equation}
\left| \Psi _{m}\right\rangle \equiv \left| \Psi \right\rangle
_{a_{m}1_{m}...N_{m}}=\frac{1}{\sqrt{2}}\left( \left| 00...0\right\rangle
_{a_{m}1_{m}...N_{m}}+\left| 11...1\right\rangle
_{a_{m}1_{m}...N_{m}}\right)   \label{e1}
\end{equation}
of which qubits $a_{m}$ are with Alice and qubits $n_{m}$ with Bob $n.$

\item[a2)]  For a given $m,$ Alice measures her qubit $a_{m}$ in the basis $%
\mathcal{B}_{z}=\{\left| 0\right\rangle ,\left| 1\right\rangle \},$ then
asks Bobs to do so with their qubits $n_{m}.$ All the parties obtain the
same outcome $j_{m}^{z}$ where $j_{m}^{z}=0$ $(j_{m}^{z}=1)$ if they find $%
\left| 0\right\rangle $ $(\left| 1\right\rangle ).$

\item[a3)]  Alice publicly broadcasts the value $x_{m}=q_{m}\oplus j_{m}^{z}$
$(\oplus $ denotes an addition mod 2).

\item[a4)]  Each Bob decodes Alice's secret bit as $q_{m}=x_{m}\oplus
j_{m}^{z}.$
\end{enumerate}

This problem-giving process is absolutely secure because $j_{m}^{z},$ for
each $m,$ takes on the value of either $0$ or $1$ with an equal probability
resulting in a truly random string $\{j_{m}^{z}\}$ which Alice uses as a
one-time-pad to encode her secret problem $\{q_{m}\}$ simultaneously for all
Bobs who also use $\{j_{m}^{z}\}$ to decode Alice's problem.

\vskip 0.5cm

\noindent \textbf{The solution-collecting process}

After a predetermined period of time depending on the problem difficulty
level Alice collects the solution from independent Bobs as follows.

\begin{enumerate}
\item[b1)]  Alice and Bobs share beforehand a large number of ordered
nonidentical $(N+1)$-partite GHZ states in the form 
\begin{equation}
\left| \Phi _{m}\right\rangle \equiv \left| \Phi \right\rangle
_{a_{m}1_{m}...N_{m}}=U_{m}\left| \Psi \right\rangle _{a_{m}1_{m}...N_{m}}
\label{e2}
\end{equation}
with 
\begin{equation}
U_{m}=I_{a_{m}}\otimes u(s_{1_{m}})\otimes u(s_{2_{m}})\otimes ...\otimes
u(s_{N_{m}})  \label{U}
\end{equation}
where $I_{a_{m}}$ is the identity operator acting on qubit $a_{m}$ and 
\begin{equation}
u(s_{n_{m}})=(\left| 0\right\rangle \left\langle 1\right| +\left|
1\right\rangle \left\langle 0\right| )^{s_{n_{m}}}
\label{u}
\end{equation}
is a unitary operator
acting on qubit $n_{m}.$ For each $n$ and $m,$ the value of $s_{n_{m}}$
chosen at random between $0$ and $1$ is known only to Alice but by no means
to any other person including Bobs. Qubits $a_{m}$ are with Alice and qubits 
$n_{m}$ with Bob $n.$

\item[b2)]  For a given $m,$ Alice measures her qubit $a_{m}$ in $\mathcal{B}%
_{z}$ with the outcome $j_{a_{m}}^{z}=\{0,1\},$ then asks Bobs to do so with
their qubits $n_{m}$ with the outcome $j_{n_{m}}^{z}=\{0,1\}.$

\item[b3)]  Each Bob $n$ publicly broadcasts the value $y_{nm}=r_{nm}\oplus
j_{n_{m}}^{z}.$

\item[b4)]  Alice decodes the solution of Bob $n$ as $r_{nm}=y_{nm}\oplus
\left[ \delta _{0,s_{n_{m}}}j_{a_{m}}^{z}+\delta
_{1,s_{n_{m}}}(j_{a_{m}}^{z}\oplus 1)\right] .$
\end{enumerate}

In the solution-collecting process the outcomes $j_{a_{m}}^{z}$ and $%
j_{n_{m}}^{z}$ are not the same anymore in general, but they are dynamically
correlated as $j_{n_{m}}^{z}=\delta _{0,s_{n_{m}}}j_{a_{m}}^{z}+\delta
_{1,s_{n_{m}}}(j_{a_{m}}^{z}\oplus 1).$ This correlation allows only Alice
who knows the value of $\{s_{n_{m}}\}$ to decode the solution of a Bob after
she obtains her own measurement outcome $j_{a_{m}}^{z}.$ As is clear, each of 
the $N$ strings $\{j_{1_{m}}^{z}\},$ $\{j_{2_{m}}^{z}\},$ 
$...,\{j_{N_{m}}^{z}\}$ appears truly random and each such a string is used by
a Bob and Alice only one time to encode/decode a secret solution $\{r_{nm}\}.
$ The above solution-collecting process is therefore absolutely secure as
well.

The essential condition to ensure absolute security of the quantum exam is a
prior sharing of the entangled states $\{\left| \Psi _{m}\right\rangle \}$
and $\{\left| \Phi _{m}\right\rangle \}$ between the teacher Alice and the
students Bobs. It is therefore necessary to propose methods for multipartite
entanglement sharing.

\vskip 0.5cm

\noindent \textbf{The }$\left| \Psi _{m}\right\rangle $-\textbf{sharing
process}

Alice and Bobs can securely share the states $\{\left| \Psi
_{m}\right\rangle \}$ as follows.

\begin{enumerate}
\item[c1)]  Alice generates a large enough number of identical states $%
\left| \Psi _{m}\right\rangle $ defined in Eq. (\ref{e1}) \cite{n1}. For
each such state she keeps qubit $a_{m}$ and sends qubits $1_{m},$ $2_{m},$ $%
...,$ $N_{m}$ to Bob $1,$ Bob $2,$ $...,$ Bob $N,$ respectively. Before
sending a qubit $n_{m}$ Alice authenticates Bob $n$ of that action.

\item[c2)]  After receiving a qubit each Bob also authenticates Alice
independently.

\item[c3)]  Alice selects at random a subset $\{\left| \Psi
_{l}\right\rangle \}$ out of the shared $\left| \Psi _{m}\right\rangle $%
-states and lets Bobs know that subset. For each state of the subset Alice
measures her qubit randomly in $\mathcal{B}_{z}$ or in $\mathcal{B}%
_{x}=\{\left| +\right\rangle ,\left| -\right\rangle \}$ with $\left| \pm
\right\rangle =(\left| 0\right\rangle \pm \left| 1\right\rangle )/\sqrt{2}\}.
$ Then she asks every Bob to measure their qubits in the same basis as hers.
Alice's (Bobs') outcome in $\mathcal{B}_{z}$ is $%
j_{a_{l}}^{z}(j_{n_{l}}^{z})=\{0,1\}$ corresponding to finding $\{\left|
0\right\rangle ,\left| 1\right\rangle \}$ and that in $\mathcal{B}_{x}$ is $%
j_{a_{l}}^{x}(j_{n_{l}}^{x})=\{+1,-1\}$ corresponding to finding $\{\left|
+\right\rangle ,\left| -\right\rangle \}.$

\item[c4)]  Alice requires each Bob to publicly reveal the outcome of each
his measurement and makes an analysis. For those measurements in $\mathcal{B}%
_{z}$ she compares $j_{a_{l}}^{z}$ with $j_{n_{l}}^{z}:$ if $%
j_{a_{l}}^{z}=j_{n_{l}}^{z}$ $\forall n$ it is all-right, otherwise she
realizes a possible attack of an outsider Eve in the quantum channel. As for
measurements in $\mathcal{B}_{x}$ she compares $j_{a_{l}}^{x}$ with $%
J_{l}^{x}=\prod_{n=1}^{N}j_{n_{l}}^{x}:$ if $j_{a_{l}}^{x}=J_{l}^{x}$ it is
all-right \cite{n2}, otherwise there is Eve in the line. If the error rate
exceeds a predetermined small value Alice tells Bobs to restart the whole
process, otherwise they record the order of the remaining shared $\left|
\Psi _{m}\right\rangle $-states and can use them for the problem-giving
process following the steps from a1) to a4).
\end{enumerate}

\noindent \textbf{The }$\left| \Phi _{m}\right\rangle $-\textbf{sharing
process}

The states $\{\left|\Phi_m\right\rangle \}$ cab be securely shared between 
the participants as follows.

\begin{enumerate}
\item[d1)]  Alice generates a large enough number of identical states $%
\{\left| \Psi _{m}\right\rangle \}$ \cite{n1}. She then applies on each of
the identical states a unitary operator $U_{p}$ determined by Eq. (\ref{U})
to transform them into the $\left| \Phi _{p}\right\rangle $-states defined
in Eq. (\ref{e2}) which are nonidentical states \cite{n3}. Afterward, for
each $\left| \Phi _{p}\right\rangle ,$ she keeps qubit $a_{p}$ and sends
qubits $1_{p},$ $2_{p},$ $...,$ $N_{p}$ to Bob $1,$ Bob $2,$ $...,$ Bob $N,$
respectively. Before sending a qubit $n_{p}$ Alice authenticates Bob $n$ of
that action.

\item[d2)]  After receiving a qubit each Bob also authenticates Alice
independently.

\item[d3)]  Alice selects at random a large enough subset $\{\left| \Phi
_{l}\right\rangle \}$ out of the shared $\left| \Phi _{p}\right\rangle $%
-states and lets Bobs know that subset. For each state $\left| \Phi
_{l}\right\rangle $ of the subset Alice measures her qubit randomly in
either $\mathcal{B}_{z}$ or $\mathcal{B}_{x},$ then asks Bobs to measure
their qubits in the same basis as hers.

\item[d4)]  Alice requires each Bob to publicly reveal the outcome of each
his measurement and makes a proper analysis. For those measurements in $%
\mathcal{B}_{z}$ she verifies the equalities $j_{a_{l}}^{z}=\delta
_{0,s_{n_{l}}}j_{n_{l}}^{z}+\delta _{1,s_{n_{l}}}(j_{n_{l}}^{z}\oplus 1).$
If the equalities hold for every $n$ it is all-right, otherwise the quantum
channel was attacked. As for measurements in $\mathcal{B}_{x}$ she compares $%
j_{a_{l}}^{x}$ with $J_{l}^{x}=\prod_{n=1}^{N}j_{n_{l}}^{x}:$ if $%
j_{a_{l}}^{x}=J_{l}^{x}$ it is all-right \cite{n4}, otherwise the quantum
channel was attacked. If the error rate exceeds a predetermined value Alice
tells Bobs to restart the whole process, otherwise they record the order of
the remaining shared $\left| \Phi _{p}\right\rangle $-states and can use
them for the solution-collecting process following the steps from b1) to b4).
\end{enumerate}

\noindent \textbf{Security of the entanglement-sharing process}

To gain useful information about the exam, Eve must attack the quantum
channel during the entanglement-sharing process. Below are several types of
attack that Eve commonly uses.

\textit{Measure-Resend Attack}. In $\mathcal{B}_{z}$ Eve measures the qubits
emerging from Alice and then resends them on to Bobs. After Eve's
measurement the entangled state collapses into a product state and her
attack is detectable when Alice and Bobs use $\mathcal{B}_{x}$ for a
security check \cite{n5}.

\textit{Disturbance Attack}. If Alice and Bobs check security only by
measurement outcomes in $\mathcal{B}_{x},$ then Eve, though cannot gain any
information, is able to make the protocol to be denial-of-service. Namely,
for each $n,$ on the way from Alice to Bob $n,$ Eve applies on qubit $n$ an
operator $u(v_{n_{m}})$ as defined in Eq. (\ref{u})
with $v_{n_{m}}$ randomly taken as either $0$ or $1,$
then lets the qubit go on its way. By doing so the disturbed states become
truly random and totally unknown to everybody, hence no cryptography is possible at all.
Though measurements in $\mathcal{B}_{x}$ cannot detect this type of attack 
\cite{n6}, those in $\mathcal{B}_{z}$ can \cite{n7}.

\textit{Entangle-Measure Attack}. Eve may steal some information by
entangling her ancilla (prepared, say, in the state $\left| \chi
\right\rangle _{E})$ with a qubit $n$ (assumed to be in the state $\left|
i\right\rangle _{n})$ before the qubit reaches Bob $n:$ $\left| \chi
\right\rangle _{E}\left| i\right\rangle _{n}\rightarrow \alpha \left| \chi
_{i}\right\rangle _{E}\left| i\right\rangle _{n}+\beta \left| \overline{\chi
_{i}}\right\rangle _{E}\left| i\oplus 1\right\rangle _{n}$ where $|\alpha
|^{2}+|\beta |^{2}=1$ and $_{E}\left\langle \chi _{i}\right. \left| 
\overline{\chi _{i}}\right\rangle _{E}=0.$ After Bob $n$ measures his qubit
Eve does so with her ancilla and thus can learn about the Bob's outcome.
Yet, with a probability of $|\beta |^{2}$ Eve finds $\left| \overline{\chi
_{i}}\right\rangle _{E}$ in which case she is detected if the security check
by Alice and Bobs is performed in $\mathcal{B}_{z}$ \cite{n8}.

\textit{Intercept-Resend Attack}. Eve may create her own entangled states $%
\left| \Psi ^{\prime }\right\rangle _{a_{m}^{\prime }1_{m}^{\prime
}...N_{m}^{\prime }}$ $(\left| \Phi ^{\prime }\right\rangle _{a_{m}^{\prime
}1_{m}^{\prime }...N_{m}^{\prime }}=U_{m}^{\prime }\left| \Psi ^{\prime
}\right\rangle _{a_{m}^{\prime }1_{m}^{\prime }...N_{m}^{\prime }}$ where $%
U_{m}^{\prime }=I_{a_{m}^{\prime }}\otimes u(s_{1_{m}}^{\prime })\otimes
u(s_{2_{m}}^{\prime })\otimes ...\otimes u(s_{N_{m}}^{\prime })$ with $%
\{s_{n_{m}}^{\prime }\}$ an arbitrary random string). Then she keeps qubit $a_{m}^{\prime }
$ and sends qubit $n_{m}^{\prime }$ to Bob $n.$ When Alice sends qubits $%
n_{m}$ to Bobs Eve captures and stores all of them. Subsequently, after
Alice's and Bobs' measurements, Eve also measures her qubits $a_{m}^{\prime }
$ and the qubits $n_{m}$ she has kept to learn the corresponding keys. This
attack is detected as well when Alice and Bobs use $\mathcal{B}_{z}$%
-measurement outcomes for their security-check \cite{n9}.

\textit{Masquerading Attack}. Eve may pretend to be a Bob in the $\left|
\Psi _{m}\right\rangle $-sharing process to later obtain Alice's problem.
Likewise, she may pretend to be Alice in the $\left| \Phi _{m}^{\prime
}\right\rangle $-sharing process to later collect Bobs' solutions. Such
pretenses are excluded because each Bob after receiving a qubit has to
inform Alice and Alice before sending a qubit has also to inform all Bobs.
The classical communication channels Alice and Bobs possess have been
assumed highly authentic so that any disguisement must be disclosed.

\vskip 0.5cm

\noindent \textbf{2.2. Asymptotically secure protocol}

In some circumstances an urgent exam needs to be organized but no prior
quantum nonlocal arrangements are available at all. We now propose a
protocol to directly accomplish such an urgent task. At that aim, Alice has
to have at hand a large number of states $\{\left| \Psi _{m}\right\rangle \}$
and $\{\left| \Phi _{m}\right\rangle =U_{m}\left| \Psi _{m}\right\rangle \}.$
Let $M$ $(M^{\prime })$ be length of Alice's problem (Bobs' solution) and $T$
the time provided for Bobs to solve the problem.
\vskip 0.5cm
\noindent \textbf{The direct problem-giving process}

Alice can directly give her problem to Bobs by ``running'' the following
program.

\begin{enumerate}
\item[e1)]  $m=0.$

\item[e2)]  $m=m+1.$ Alice picks up a state $\left| \Psi _{m}\right\rangle ,$
keeps qubit $a_{m}$ and sends qubits $1_{m},$ $2_{m},$ $...,$ $N_{m}$ to Bob 
$1,$ Bob $2,$ $...,$ Bob $N,$ respectively. Before doing so Alice informs
all Bobs via her authentic classical channels.

\item[e3)]  Each Bob confirms receipt of a qubit via their authentic
classical channels.

\item[e4)]  Alice switches between two operating modes: the control mode
(CM) with rate $c$ and the message mode (MM) with rate $1-c.$ Alice lets
Bobs know which operating mode she chose.

\begin{enumerate}
\item[e4.1)]  If CM is chosen, Alice measures qubit $a_{m}$ randomly in $%
\mathcal{B}_{z}$ or $\mathcal{B}_{x}$ with an outcome $j_{a_{m}}^{z}$ or $%
j_{a_{m}}^{x},$ then lets Bobs know her basis choice and, asks them to
measure their qubits $n_{m}$ in the chosen basis. After measurements each
Bob publicly publishes his outcome $j_{n_{m}}^{z}$ or $j_{n_{m}}^{x}.$ Alice
analyzes the outcomes: if $%
j_{a_{m}}^{z}=j_{1_{m}}^{z}=j_{2_{m}}^{z}=...=j_{N_{m}}^{z}$ or $%
j_{a_{m}}^{x}=\prod_{n=1}^{N}j_{n_{m}}^{x}$ she sets $m=m-1$ and goes to
step e2) to continue, else she tells Bobs to reinitialize from the beginning
by going to step e1).

\item[e4.2)]  If MM is chosen, Alice measures qubit $a_{m}$ in $\mathcal{B}%
_{z}$ with an outcome $j_{a_{m}}^{z}$ and publicly reveals $%
x_{m}=j_{a_{m}}^{z}\oplus q_{m}.$ Each Bob measures his qubit also in $%
\mathcal{B}_{z}$ with an outcome $j_{n_{m}}^{z},$ then decodes Alice's
secret bit as $q_{m}=j_{n_{m}}^{z}\oplus x_{m}.$ If $m<M$ Alice goes to step
e2) to continue, else she publicly announces: \textit{``My problem has been
transferred successfully to all of you. Please return your solution after
time }$T".$
\end{enumerate}
\end{enumerate}

\noindent \textbf{The direct solution-collecting process}

After time $T$ Alice can directly collect Bobs' solutions by ``running''
another program as follows.

\begin{enumerate}
\item[g1)]  $m=0.$

\item[g2)]  $m=m+1.$ Alice picks up a $\left| \Phi _{m}\right\rangle ,$
keeps qubit $a_{m}$ and sends qubits $1_{m},$ $2_{m},$ $...,$ $N_{m}$ to Bob 
$1,$ Bob $2,$ $...,$ Bob $N,$ respectively. Before doing so Alice informs
all Bobs via her authentic classical channels.

\item[g3)]  Each Bob confirms receipt of a qubit via their authentic
classical channels.

\item[g4)]  Alice switches between two operating modes: the CM with rate $c$
and the MM with rate $1-c.$ Alice lets Bobs know which operating mode she
chose.

\begin{itemize}
\item[g4.1)]  If CM is chosen, Alice measures qubit $a_{m}$ randomly in $%
\mathcal{B}_{z}$ or $\mathcal{B}_{x}$ with an outcome $j_{a_{m}}^{z}$ or $%
j_{a_{m}}^{x},$ then lets Bobs know her basis choice and, asks them to
measure their qubits $n_{m}$ in the chosen basis. After measurements each
Bob publicly publishes his outcome $j_{n_{m}}^{z}$ or $j_{n_{m}}^{x}.$ Alice
analyzes the outcomes: if $j_{a_{m}}^{z}=\delta
_{0,s_{n_{m}}}j_{n_{m}}^{z}+\delta _{1,s_{n_{m}}}(j_{n_{m}}^{z}\oplus 1)$
for every $n$ or $j_{a_{m}}^{x}=\prod_{n=1}^{N}j_{n_{m}}^{x}$ she sets $m=m-1
$ and goes to step g2) to continue, else she tells Bobs to reinitialize from
the beginning by going to step g1).

\item[g4.2)]  If MM is chosen, Alice measures qubit $a_{m}$ in $\mathcal{B}%
_{z}$ with an outcome $j_{a_{m}}^{z}$ and each Bob measures his qubit also
in $\mathcal{B}_{z}$ with an outcome $j_{n_{m}}^{z}.$ Each Bob publicly
reveals $y_{nm}=r_{nm}\oplus j_{n_{m}}^{z}$ and Alice decodes Bobs' secret
bits as $r_{nm}=y_{nm}\oplus \left[ \delta
_{0,s_{n_{m}}}j_{a_{m}}^{z}+\delta _{1,s_{n_{m}}}(j_{a_{m}}^{z}\oplus
1)\right] $ for $n=1,2,...,N.$ If $m<M^{\prime }$ Alice goes to step g2) to
continue, else she publicly announces: \textit{``Your solutions have been
collected successfully''}.
\end{itemize}
\end{enumerate}

As described above, in the direct problem-giving (solution-collecting)
process Alice alternatively gives (collects) secret bits and checks Eve's
eavesdropping. These direct protocols also stand against the types of attack
mentioned above. The protocols terminate immediately whenever Eve is
detected in a control mode. However, Eve might get a partial information
before her tampering is disclosed. Such an information leakage can be
reduced as much as Alice wants by increasing the control mode rate $c$ at
the expense of reducing the information transmission rate $r=1-c.$ For short
strings $Q$ and $R_{n}$ (see Eq. (\ref{Q}) and Eq. (\ref{R})) Eve's
detection probability may be quite small. But, the longer the strings the
higher the detection probability. In the long-string limit the detection
probability approaches one, i.e. Eve is inevitably detected. In this sense,
the direct quantum exam protocols are asymptotically secure only.
\vskip 0.5cm
\noindent \textbf{3. Conclusion}

We have proposed two protocols for organizing a quantum exam \cite{n10}
basing on a judicious use of appropriate multipartite entangled states. The
first protocol is absolutely secure iff the participants have successfully
shared the necessary entanglement in advance. We also provide methods for
sharing the multipartite entanglement in the presence of a potential
eavesdropping outsider. The second protocol can be processed directly
without a prior entanglement sharing. This advantage is however compromised
by a lower confidentiality level or by a slower information transmission
rate. Both the protocols have been shown to sustain various kinds of attacks
such as measure-resend attack, disturbance attack, entangle-measure attack,
intercept-resend attack and masquerading attack. Our protocols work well in
an idealized situation with perfect entanglement sources/measuring devices
and in noiseless quantum channels which we have assumed for simplicity. We
are planning to further develop our protocols to cope with more realistic
situations.

\vskip 0.5cm

\noindent \textbf{Acknowledgments.} 

The author is grateful to Professor
Hai-Woong Lee from KAIST for useful discussion and comments. This research
was supported by a Grant (TRQCQ) from the Ministry of Science and Technology
of Korea and also by a KIAS R\&D Fund No 6G014904.


\begin{thebibliography}{99}
\bibitem{n1}  A state $\left| \Psi _{m}\right\rangle \equiv \left| \Psi
\right\rangle _{a_{m}1_{m}...N_{m}}$ can be generated from the product state 
$\left| 00...0\right\rangle _{a_{m}1_{m}...N_{m}}$ as $\left| \Psi
\right\rangle _{a_{m}1_{m}...N_{m}}=CNOT_{a_{m}N_{m}}\otimes ...\otimes
CNOT_{a_{m}2_{m}}\otimes CNOT_{a_{m}1_{m}}\otimes H_{a_{m}}\left|
00...0\right\rangle _{a_{m}1_{m}...N_{m}},$ where $CNOT_{xy}$ is a
control-NOT gate with $x$ $(y)$ being the control (target) qubit and $H_{x}$
is the Hadamard gate acting on qubit $x.$
\bibitem{n2}  This checking strategy comes out from the fact that a state $%
\left| \Psi \right\rangle _{a_{l}1_{l}...N_{l}}$ always has a positive
parity, i.e. $j_{a_{l}}^{x}\prod_{n=1}^{N}j_{n_{l}}^{x}=+1$ definitely.
\bibitem{n3}  Since each $U_{p}$ is managed by Alice alone, she is the only
one who knows the parameters $s_{1_{p}},$ $s_{2_{p}},$ $...,$ $s_{N_{p}}$
though each of them is randomly chosen between $0$ and $1.$ That is, only
Alice is able to distinguish states $\left| \Phi _{p}\right\rangle $ with
certainty.
\bibitem{n4}  This checking strategy comes out from the fact that a state $%
\left| \Phi \right\rangle _{a_{l}1_{l}...N_{l}}$ also has definitely a
positive parity.
\bibitem{n5}  This is because any product state $\left|
i_{a}i_{1}i_{2}...i_{N}\right\rangle _{a1...N}$ with $i_{a},i_{n}\in \{0,1\},
$ when measured in $\mathcal{B}_{x},$ yields either ``plus'' or ``minus''
parity with an equal probability.
\bibitem{n6}  Because any entangled state $\left( \left|
i_{a}i_{1}i_{2}...i_{N}\right\rangle _{a1...N}+\left| \overline{i_{a}}%
\overline{i_{1}}\overline{i_{2}}...\overline{i_{N}}\right\rangle
_{a1...N}\right) /\sqrt{2},$ with $\overline{i}$ the negation of $i=\{0,1\},$
has definitely a positive parity.
\bibitem{n8}  Because in this case what Bob actually finds is $%
\left| i\oplus 1\right\rangle _{n}$ instead of the should-be $\left|
i\right\rangle _{n}.$
\bibitem{n9}  Because there are no correlations at all between qubit 
$a_{m}$ and qubits $n_{m}^{\prime }.$
\bibitem{n10}  If one likes, the name ``quantum exam'' can alternatively be
replaced by ``quantum multiparty interview'' (or ``quantum ballot'') for
which Alice plays the role of the interviewer (or the selection committee
chair) and Bobs play the role of interviewees (or balloters).
\end{thebibliography}
\end{document}